# Dense $e^-e^+$ plasma formation in magnetic dipole wave: vacuum breakdown by 10-PW class lasers

A.V. Bashinov[1*], E.S. Efimenko[1], A.A. Muraviev[1], V.D. Volokitin[2], I.B. Meyerov[2], G. Leuchs[1,3], A.M. Sergeev[1], A.V. Kim[1]

[1]*Institute of Applied Physics, Russian Academy of Sciences, Nizhny Novgorod 603950, Russia*
[2]*Lobachevsky State University of Nizhni Novgorod, Nizhny Novgorod 603950, Russia*
[3]*Max Planck Institute for the Science of Light, Erlangen, Germany*

When studying the interaction of matter with extreme fields using multipetawatt lasers, there are two limiting cases maximizing either the electric field or the magnetic field. Here, the main attention is paid to the optimal configuration of laser beams in the form of an m-dipole wave, which maximizes the magnetic field, and the corresponding production of pair plasma via a QED cascade using 10-PW class lasers. We show that the threshold of vacuum breakdown with respect to avalanche-like pair generation is about 10 PW. Using 3D PIC modeling in the specified fields, we go deeper into the physics of vacuum breakdown, i.e. we examined in detail the individual trajectories of particles produced in inhomogeneous electric and magnetic fields, the space-time distributions of pair densities on the avalanche stage, and the energy distributions of charged particles and gamma photons. The forming plasma structures represent concentric rings around the central magnetic axis, which can result in significant change of laser-plasma interaction in comparison with the case of an e-dipole wave.

## 1 Introduction

Currently, actively developing multipetawatt laser facilities [1] can become a unique tool for studying the properties of quantum vacuum and quantum electrodynamics (QED) processes in extremely strong fields, as well as for modeling astrophysical phenomena in laboratory conditions [2, 3]. Much attention has been paid to the optimal configurations of laser beams with respect to maximizing the electric field, assuming its leading role in establishing the main physical effects. However, of course, there are many fundamental effects in which the magnetic field is a key factor, in particular, in such an important problem as QED-creation of electron-positron pairs by ultra-high-power lasers [4], both electric and magnetic fields are equally important; an electron is accelerated to high energy in the driving electric field, but it emits gamma photons mainly on the curved part of the trajectory induced by a strong magnetic field. The gamma photon, in turn, decays in strong fields into a pair of an electron and a positron, which, in turn, can get accelerated to high energies. Such a repetitive process is usually called a QED cascade [5], which, under certain conditions, can lead to an avalanche-like growth of the number of pairs, thereby initiating a vacuum breakdown [6-8]. Ultimately, a dense pair plasma can be created, in which the plasma structures and their properties substantially depend on the configuration of the electric and magnetic fields [9-21]. As shown in the limiting case of a multi-beam setup in the form of a converging e-dipole wave [22, 23], when the focal volume is minimal and the electric field is highest, extreme plasma states can be created, paving the way to the quantum pair plasma and achieving the Schwinger field [24, 25].

At present, we are interested in the case of a multi-beam laser setup that maximizes the magnetic field, and we begin our consideration with the limiting case in the form of a converging m-dipole wave. In this work, the main emphasis is placed on the physics of the effect of vacuum breakdown by ultrahigh-power laser fields, i.e., what parameters of the laser we need and what we expect when a vacuum breakdown occurs. We show that the forming plasma structures represent concentric rings around the central magnetic axis, which can result in significant change of laser-plasma interaction in comparison with the case of an e-dipole wave.

This paper, which is mainly based on 3D modeling using the QED-PIC PICADOR code [26, 27], is organized as follows. In Sec. 2, we briefly describe the distributions of the electric and magnetic fields of a standing m-dipole wave, which is important for understanding the individual dynamics of particles. In Sec. 3, we show the macrocharacteristics of vacuum breakdown. In Sec. 4, we discuss in detail the individual trajectories of charged particles and gamma photons emitted in a given field. These results illustrate the interplay between various effects in non-homogeneous laser fields. The vacuum breakdown threshold is predicted in Sec. 5. In Sec.6, the possibilities of the experimental approach are discussed. Finally, in Sec. 7 the conclusions are summarized.

## 2 Field structure of the m-dipole wave

In order to get an insight into the development of vacuum breakdown we consider the ideal case of the multi-beam configuration of the laser setup in the form of an incoming m-dipole wave [22]. As soon as the incident m-dipole wave reaches the focus a diverging wave appears. As a result, particle motion and development of the QED cascade most of time occur in a standing m-dipole wave. The exact analytical expressions for the standing fields are as follows [22]:

$$\vec{E} = F_0\sqrt{P}\cos(\omega t)\frac{\rho}{R}\left[\frac{\sin(kR)}{(kR)^2} - \frac{\cos(kR)}{kR}\right]\vec{e_\varphi}$$

$$\vec{B} = -F_0\sqrt{P}\sin(\omega t)\left\{\left[\frac{\sin(kR)}{kR}\left(1 - \frac{1+(kz)^2}{(kR)^2} + \frac{3(kz)^2}{(kR)^4}\right) + \frac{\cos(kR)}{(kR)^2}\left(1 - \frac{3(kz)^2}{(kR)^2}\right)\right]\vec{e_z} + \frac{\rho z}{R^2}\left[\frac{\sin(kR)}{kR}\left(\frac{3}{(kR)^2} - 1\right) - \frac{3\cos(kR)}{(kR)^2}\right]\vec{e_\rho}\right\}$$

where $R = \sqrt{\rho^2 + z^2}$, $\rho, z$ are radial and axial coordinates in the cylindrical coordinate system, $P$ is the wave power in petawatts, $F_0 = \frac{2e}{mc^2}\sqrt{\frac{3}{c}}10^{22}\,\mathrm{erg}\cdot\mathrm{s}^{-1} \approx 1174$, $e$ is the elementary charge, $m$ is the positron mass, $c$ is the light velocity, $t$ is time, $\omega \approx 2.1 \times 10^{15}\,\mathrm{s}^{-1}$ is the wave frequency corresponding to the wave period $T = 3$ fs and the wavelength $\lambda = 0.9$ μm, $k = \omega/c$ is the wave number, $\vec{e}_{e_{\rho,\varphi,z}}$ are radial, azimuthal and axial unit vectors. Fields are normalized to the relativistic field value $mc\omega/e$.

The magnetic field has a poloidal structure and the electric field is toroidal (see Figure 1 (a)). Both electric and magnetic fields are axially symmetric. In the coordinate system we have chosen, the magnetic field has radial and axial components while the electric field only has the azimuthal component. In the focal spot the magnetic field is directed predominantly along the $z$ axis. The characteristic extrema are in Figure 1 (b). The maximum amplitude of the magnetic field $a_B = a = 2F_0\sqrt{P}/3 \approx 780\sqrt{P}$ is achieved at the point $\rho = 0, z = 0$, which we will further refer to as the central point or the center of the wave. The maximum amplitude of the electric field $a_E = 0.65a \approx 510\sqrt{P}$ is achieved on a circle lying in the central plane $z = 0$ with $\rho = 0.33\lambda$. Near the central point electric and magnetic fields can be approximated as

$$\vec{E} = akr\left(\frac{1}{2} - \frac{(kz)^2}{20}\right)\vec{e_\varphi}$$
$$\vec{B} = a\left\{\left[\left(1 - \frac{(kz)^2}{10}\right) + (kr)^2\left(\frac{(kz)^2}{70} - \frac{1}{5}\right)\right]\vec{e_z} + \frac{k^2zr}{10}\vec{e_\rho}\right\}.$$

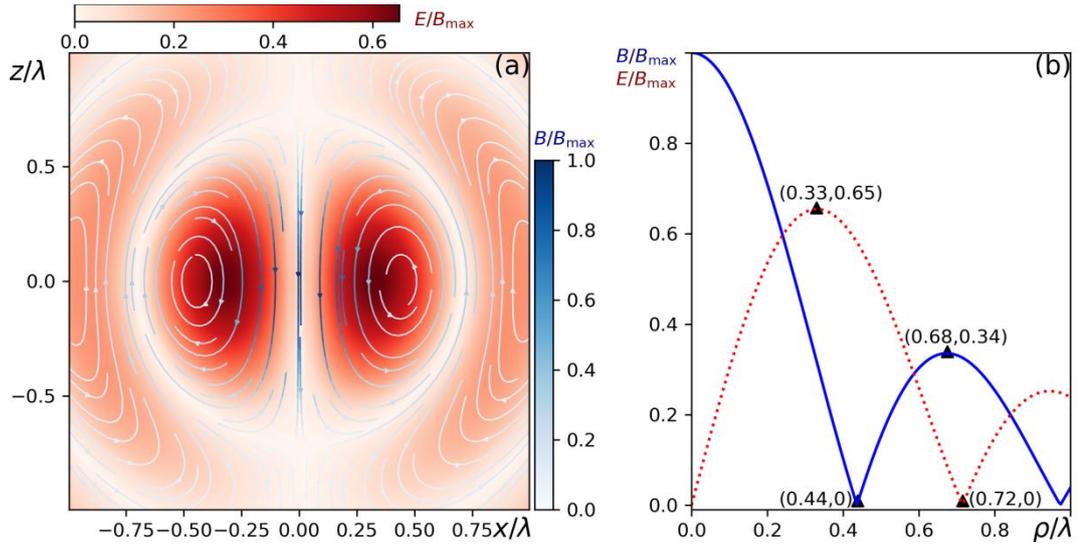

**Figure 1.** Spatial distribution of electric and magnetic fields of the standing m-dipole wave. For clarity the figure (a) is presented in coordinates $z$ and $x$ in the Cartesian coordinate system. (b) Electric and magnetic fields as function of $\rho$ in the plane z=0.

The $z$-axis corresponds to the first node of the electric field. The second node is approximately a sphere with radius $r = 0.72\lambda$. Note, that the second antinode of the magnetic field with the amplitude $0.34a_B$ is at $\rho = 0.68\lambda$.

## 3 Vacuum breakdown in the m-dipole wave

Vacuum breakdown is a process of avalanche-like production of electron-positron pairs in laser fields when particle losses due to particle escape from the region of particle generation becomes slower than generation of new particles in this region. In the power range from 9 PW to 50 PW the most prominent process of pair creation is the QED cascade [5]. This process depends on the space-time structure of fields, particles' and photons' distributions and their spectra. Moreover it can be non-local, for an example, photons generated in one region can propagate and decay in another region. As an initial step, to understand how and where vacuum breakdown occurs in fields of the m-dipole wave we consider evolution of ensembles of electrons and positrons in fields of this wave in the chosen power range with and without QED cascade taken into account. In the case without QED cascade particles may emit photons and experience recoil, but photons are prohibited from decaying into electron-positron pairs. In this paper we pay particular attention to the breakdown stage in the conditions when plasma back reaction can be neglected.

For simulations we use the 3D QED-PIC code PICADOR [27, 26]. Initially an electron-positron sphere with the radius of $\lambda$ is located in the fields of the continuous standing m-dipole wave, and the center of the sphere matches the central point. In this section we consider the stage of vacuum breakdown when there is no back reaction of particles on fields. In order to neglect the back reaction fields are set analytically in the form of the m-dipole wave. The initial number of macroparticles of each type (electron, positron) is $4 \times 10^6$. Particle motion and QED-cascade development are simulated for $20T$. The simulation box is $3\lambda \times 3\lambda \times 3\lambda$ along $x$, $y$ and $z$ axes with cell number $192 \times 192 \times 192$.

One of the main questions from the experimental point of view is the amount of total laser power we must deliver to a seed target to trigger the vacuum breakdown. To determine the threshold power for vacuum breakdown in each simulation we calculate the avalanche growth rate $\Gamma = \ln\left(\frac{N_p(t_s+lT)}{N_p(t_s)}\right)/lT$. Here $N_p(t)$ is the number of electron-positron pairs in the simulation box at the moment of time $t$, $t_s$ is the moment of time chosen so that the exponential growth of the amount of particles becomes steady, $l$ is an integer number. It should be noted that the growth rate $\Gamma$ is the sum of the production rate of new particles and the rate of particle escape from the focal region. Very roughly $t_s \propto \Gamma^{-1}$, but it should not be less than $1.5T$, we explain the minimum boundary of $t_s$ below. $\Gamma$ does not depend on the region in which the number of particles is counted, however, the larger this region is, the longer $t_s$ must be. In simulations $t_s = 18T$ and $l = 2$. $\Gamma$ as function of power is depicted in Figure 2.

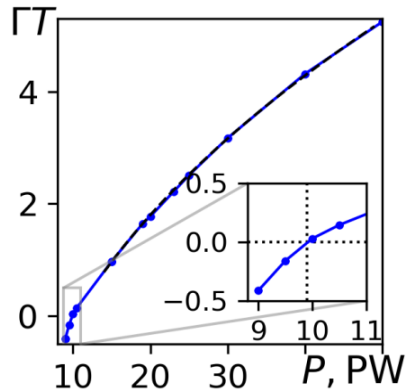

**Figure 2. Dependence of avalanche growth rate in fields of the m-dipole wave on its power. Markers correspond to the simulation result. Dashed line shows approximation.**

First, simulations show that the vacuum breakdown threshold power is $P_{th} \approx 10$ PW (an inset in Figure 2). We should also note that although the field structure is different from the e-dipole wave [22] and seems nonoptimal to trigger vacuum breakdown, nevertheless, the threshold powers are comparable: 10 PW in the case of the m-dipole wave versus 7.2 PW in the case of the e-dipole wave [28].

Second, laser fields with an m-dipole structure with power about 10PW can produce dense electron-positron plasma in the matter of femtoseconds. For example, 2.5 wave periods are required to increase the electron-positron plasma density by two orders of magnitude in fields of a 20 PW m-dipole wave. For $P > 10$ PW an accurate approximation of the avalanche growth rate is $\Gamma T = 2.2(P^{0.4} - 2.6)$ (black dashed line in Figure 2).

For the power range when the vacuum breakdown occurs distributions of positrons at different moments of time are shown in Figure 3. These figures show the distributions at the moments of time when particles are closest to the center (see Figure 3(e)) and farthest from it (see Figure 3(a)). The particle structure is similar to several concentric oscillating cylindrical layers with heights much less than the laser wavelength. In the case with QED cascade taken into account the inner layer is oscillating around the first antinode of the electric field and the outer layer is oscillating close to the first electric field node at $\rho = 0.72\lambda$. These layers are shown by rectangles with dashed edges. Besides oscillating layers there are layers of escaping particles outside outer oscillating layers. The photon distribution also consists of several concentric cylindrical layers (see Figure 3 (b), (f)) and in a similar manner this distribution oscillates along the radius. The simulations show that the greater the wave power, the more distinct the inner layer is as compared to the outer one.

To distinguish the impact of QED cascade development from the impact of particle motion on vacuum breakdown we compare particle and photon distributions with and without QED cascade taken into account. Without QED cascade, particles are trapped and oscillate in the vicinity of the second electric field node, thus the particle distribution represents a spherical layer (see Figure 3 (c), (g)). At the polar caps of the sphere at the intersection with the z-axis, where both the electric and magnetic fields are weak, particles form bunches, which periodically contract and expand and gradually run out at an angle to the z-axis.

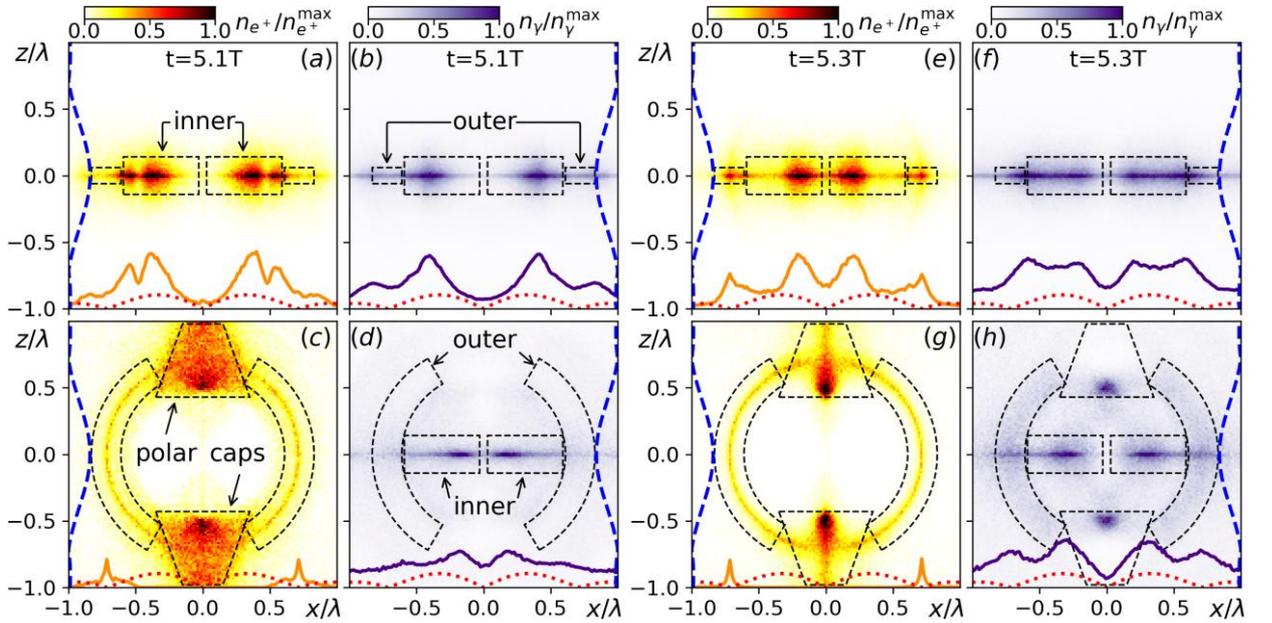

**Figure 3. Interaction of a standing m-dipole wave with seed particles at the wave power of $P = 20PW$. Positron (a),(c),(e),(g) and photon (b),(d),(f),(h) density distributions normalized to its maxima at different moments of time (a),(b),(e),(f) with and (c),(d),(g),(h) without QED cascade taken into account. The dotted red and dashed blue curves correspond to the distribution of electric and magnetic fields along x and z axes passing the origin of coordinates. Solid orange and purple curves show photon and positron distributions along the same z axis. Figures with dashed edges highlight different regions of particle and photon distributions.**

In the case without QED cascade there is only the outer layer and there is no inner layer. Particles are not trapped in a vicinity of the first electric field antinode, which means that the inner layer is the result of cascade development. Since cascade growth is faster near the $z = 0$ plane due to a slower drift from this plane along the z-axis and higher probability of photon decay, the particle distribution is more compact along the z-axis in the case when the QED cascade is taken into account. Even at the second electric field node we can observe only a small spherical sector (see Figure 3 (e)), not a large part of sphere like in Figure 3 (c), (g).

Though the particle distributions are different, photon distributions are quite similar in cases with (see Figure 3 (b), (f)) and without (see Figure 3 (d), (h)) QED cascades especially in the vicinity of the plane $z = 0$: in both cases inner and outer layers are present. Even without QED cascade there are photons propagating to the centre of the m-dipole wave from the second electric field node (a sphere with radius $0.72\lambda$), at $\rho = 0$ the photon density is nonzero, although at its local minimum. As it will be shown below, by means of these photons particles of the outer layer can supply electron-positron pairs into the inner layer.

## 4 Particle trajectories near threshold level

In contrast to conventional gas or matter optical breakdown where the ionization rate can be expressed as a function of the local value of electric field, in the case of vacuum breakdown due to QED cascade development particle motion plays a definitive role. Thus we start our consideration by systematic study of different types of trajectories in the field of a magnetic dipole wave.

A distinctive feature of particle motion in fields of the m-dipole wave is that the motion is not planar, but three dimensional. The electric field drives a particle in the azimuthal direction while the magnetic part of the Lorentz force excites radial and axial motions. In contrast, in an e-dipole wave the magnetic field is purely azimuthal and there is no force acting on a particle in azimuthal direction and motion is planar: radial and axial [23].

According to numerical simulations described above (Sec. 3) we mark two specific regions important for the QED cascade: the inner cylindrical layer in the vicinity of the first electric field antinode and the outer cylindrical layer close to the first electric field node as shown in Figure 3. The analysis of trajectories of particles from these regions is important for understanding of vacuum breakdown development.

### 4.1 Escaping particle trajectories

As follows from comparison of Figure 3 (a) and (c) in the considered range of powers particles in the inner layer do not get trapped. Most particles leave the inner layer within a time interval shorter than the wave period, although some particles may take several wave periods to escape. The closer a particle is to the *z*-axis, the longer it takes for it to leave the inner layer. In Figure 4 (a),(b) for the case of 20PW wave power we show trajectories of particles which took from 0.5T to 2.5T to escape the inner layer: each loop or U-turn corresponds to approximately 0.5T.

The escape time is determined by the radiation losses, which impede but do not stop particle escape from the inner layer. Since a particle moves in fields of the standing wave one may distinguish two alternating stages of motion: when the electric field is greater than the magnetic field and vice versa. At the first stage during which the electric field exceeds the magnetic field the curvature of the particle trajectory is very small and a particle is accelerated with negligible radiation losses and can gain energy up to $\gamma \sim a_E$. At the second stage the magnetic field becomes dominant, the curvature of particle trajectory is significantly increased which can cause strong radiation losses.

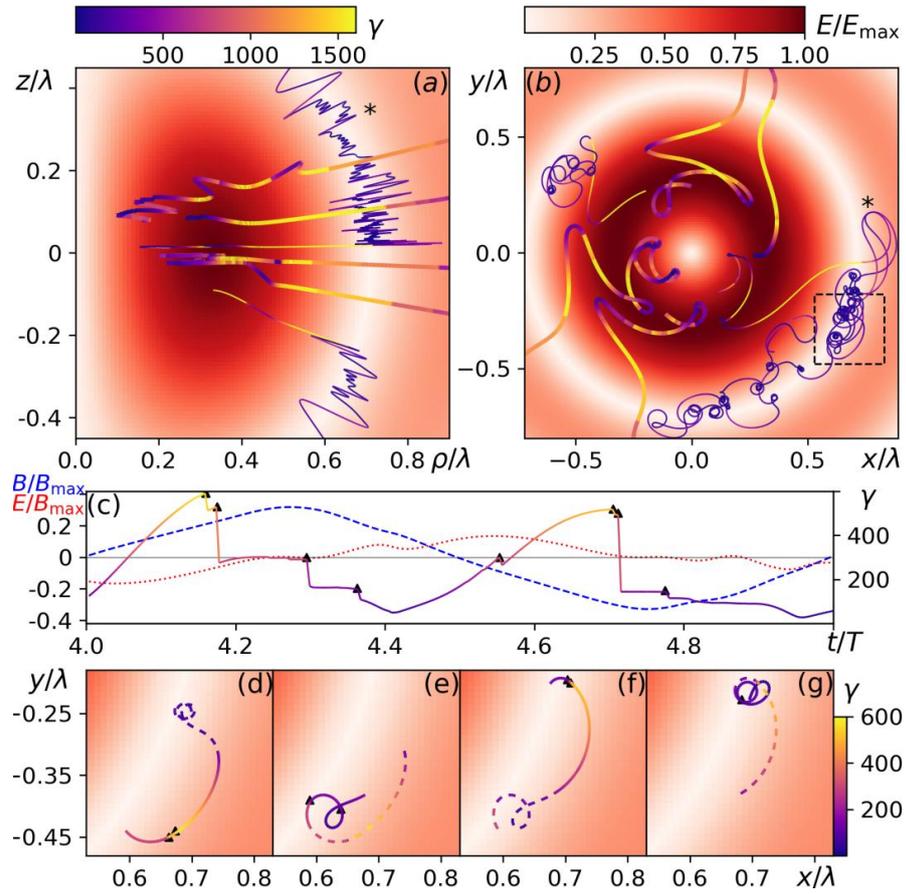

Figure 4. Positron motion in fields of the m-dipole wave. Different types of trajectories in coordinate space: (a) $z\rho$ space and (b) $xy$ space. Thin and thick lines correspond to the trapped and runaway positrons, respectively. The color along the trajectory demonstrates the particle's Lorentz-factor. Distribution of the normalized electric field amplitude is shown by gradations of red as function of (a) $z$ and $\rho$ and (b), (d)-(g) $x$ and $y$ in the plane $z = 0$. A more detailed view of positron motion close to the electric field node for a trajectory marked by asterisk in (a) and (b) is shown in (c)-(g). Solid multicolor line, solid blue line and dotted red line indicate time evolution of the positron Lorentz-factor, the z-component of the magnetic field and the azimuthal electric field along trajectory, respectively. A solid line in each subfigure (d)-(g) corresponds to the positron trajectory during quarter period in upper subfigure (c). A dashed line shows part of the trajectory during previous quarter period. Triangle marks in (c) and corresponding marks in (d)-(g) indicate moments of emission of highly energetic photons. Region considered in (d)-(g) is highlighted by a rectangle with dashed edge in (a).

If a particle loses a large enough part of its energy due to photon emission, then the Larmor radius is significantly reduced and the particle is shifted towards the electric field node but can stay in the inner region for another *0.5T*. Thus for a given particle this time is approximately a whole number of wave half-periods. The greater the wave power, the more time is needed for particles to escape. For wave powers from 9PW to 50 PW the characteristic escape time grows from 0.5T to 0.7T according to simulations.

As such, the majority of particles quickly leave the inner region. Although most of those, in turn, escape the simulation box altogether, a portion of the particles can be trapped in the outer region: in the vicinity of the following electric field nodes. Most trapped particles are trapped near the second and third electric field nodes. Below we consider trapping near the second node, which is more closely interdependent with the cascade in the inner layer through the photons emitted towards the center by trapped particles. Examples of such trajectories are shown by thin lines in Figure 4 (a), (b).

When particles leave the inner layer, they enter the region where the magnetic field dominates. If a particle keeps its energy $\gamma \sim a_E$, then it passes through this region because the maximum magnetic field in the outer layer is $0.34 a_B$ and the corresponding Larmor radius $a_E/(0.34 a_B \times k) \approx 0.3\lambda$ radius is large enough. To be trapped a particle must have a much smaller Larmor radius, which can result from energy reduction due to emission of one or several photons with a substantial part of particle's energy. From Figure 4 (a), (b) it is clear that trapped particles (thin lines) reduce their energy much more significantly than transient ones (thick lines) at the second electric field node. This trapping is the so called normal radiative trapping (NRT) [23].

When trapped, a particle shows random walking in the azimuthal direction and slow drifting along the sphere of the second electric field node to polar caps of the sphere. A particle can escape the trapped state at a distance larger than approximately $0.4\lambda$ from the $z = 0$ plane, where fields become significantly weaker and the Larmor radius increases. The escaped particle propagates at a large angle to the $z = 0$ plane, because far from this plane the radial component of the magnetic field becomes comparable with the axial component and excites large axial momentum. On average particles need approximately 3-5 $T$ to escape the trapped state in the considered range of the wave power.

In order to understand photon emission by trapped particles and their impact on the QED cascade in the inner layer we present a typical particle motion in the NRT regime during the wave period (see Figure 4 (c),(d)-(g)). The considered motion corresponds to the trajectory marked by asterisk in Figure 4 (a),(b).

During approximately the first quarter of the wave period the trapped particle is accelerated by the electric field in the vicinity of the electric field node (see Figure 4 (c),(d)). During this acceleration the growing magnetic field deflects the particle to the region with a stronger electric field, increasing the curvature of the trajectory and enhancing the emission of photons marked by triangles in Figure 4 (c), (d)-(g). This deflection can be in both directions: to the first or to the second electric field antinode. For the considered particle this results in clockwise motion in Figure 4 (d). During the first quarter of the wave period the particle makes approximately a half-turn with the maximal distance from the node of about $\Delta\rho = cT/(4\pi) \approx 0.1\lambda$. At such distance the electric field amplitude is approximately $0.4 a_E$, and thus a particle can gain a Lorentz-factor up to $0.4 a_E$. These estimates of maximal energy and particle shifts from the node are consistent with Figure 4 (a), (c).

During the second quarter period (Figure 4 (c), (e)) the magnetic field reaches its maximum. Due to several photon emissions a particle loses a major part of its energy and rotates with a small Larmor radius, i.e. remains in the trapped state.

On the next half-period (see Figure 4 (c), (f)) the z-component of the magnetic field changes its sign. As a result the particle's clockwise motion changes to counter-clockwise motion. In other respects the particle motion is similar to the first half-period. The particle is deflected to the region where the electric field stronger gaining energy again up to $\sim 0.4 a_E$. The increased magnetic field stimulates photon emission so the particle rotates, emits a number of photons, reduces its energy and Larmor radius, see Figure 4 (c), (g). Thus each half a period the particle makes a step in the azimuthal direction. This step occurs in a random direction depending on the number of rotations at the second quarter-period. The discussed properties of particle motion related to directed drift and changes of drift directions can be seen in Figure 4 (b) for the trajectory marked with asterisk.

Also Figure 4 (d)-(g) shows, that while trapped, a particle can emit photons in different directions including the direction towards the center, and energy of these photons can be comparable with particles' energy (see Figure 4 (a), (f)).

## 4.2 Energy and angular spectra of particles and photons

Based on particle motion in fields of the standing m-dipole wave below we consider characteristics of angular and energy distribution of particles. The strongest electric field with the amplitude $a_E$ is in the inner layer, where a particle before leaving this region can obtain the greatest energy of about $a_E$ (see Figure 5 (a), orange and green dashed lines). Energy of particles and photons is normalized to the electron rest energy $\varepsilon_r$. In the range of powers when the vacuum breakdown occurs the maximum value of the quantum parameter determining photon emission and photon decay is $\chi \gtrsim 1$, and a particle can lose almost all its energy in an act of photon emission [29]. Thus the maximum energy of generated photons is also close to $a_E$ (see Figure 5 (a), violet dashed line).

To analyze spectra of particles and photons which escape the focal region in simulations, their characteristics are saved when they cross the sphere with radius $1.5\lambda$. Note that the spectrum of escaping photons (see Figure 5 (a), violet solid line) has a dip in the high energy range in comparison with the spectrum of all photons in the simulation box. Photons in this energy range are most likely to decay into electron-positron pairs and a significant part of these photons cannot leave the strong field region.

A spectrum of escaping particles has distinct maxima at energy an order of magnitude less than $a_E$ (see Figure 5 (a), orange and green solid line mostly overlap). Comparing the distribution of outgoing particles to that of all particles in the simulation box there are fewer low-energy particles, since the latter are mostly trapped. Although particles can escape from the trapped state as discussed in the previous section, the majority of escaping particles originate in the inner layer. Also, the spectrum of outgoing particles has a lower high-energy tail that can be explained by energy losses due to photon emission, intensified in the vicinity of the second and the third electric field nodes.

Spatial distributions of escaping particles and photons are modulated at the double wave frequency because, as obtained in previous section, the escape time for a given particle is approximately proportional to *0.5T* and radiation losses are intensified each half of wave period. Since the cascade mainly develops in the vicinity of the $z = 0$ plane where the fields are strongest and the drift velocity in the $z$ direction is small, the angular distributions of escaping photons and particles are very narrow with a peak at polar angle $\theta = \pi/2$ and angular spread of about 1 degree (see Figure 5 (b)).

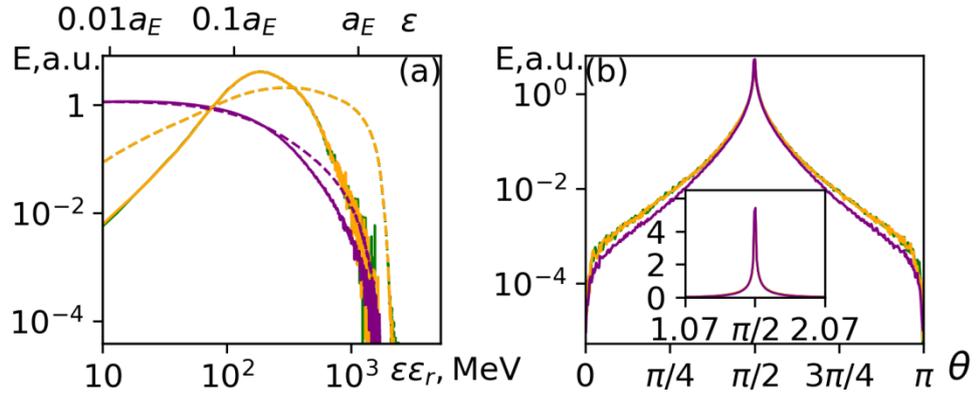

Figure 5. (a) Energy and (b) angular averaged spectra of electrons (green line), positrons (orange line) and photons (purple line) in the field of the m-dipole wave with power 20PW. Solid lines correspond to spectra of particles crossing the observation sphere of radius $r = 1.5\lambda$ averaged over the wave period. Dashed lines show averaged energy spectra in the whole simulation box. Spectra are averaged by summation of instantaneous spectra divided by the instantaneous number of particles or photons. The inset demonstrates angular distributions of particles and photons in the linear scale, lines corresponding to photons and particles overlap. Bottom and top axes in (a) show dimensional and dimensionless energy ranges, respectively, $a_E = 2285$.

Trapped particles and particles in the inner layer close to the z-axis need several wave periods to escape. During this time they drift in the *z*-direction as well. At a distance of several tenths of wavelength from the *z=0* plane the radial component of the magnetic field becomes comparable with the *z*-component. In turn, particles accelerated by the electric field in the azimuthal direction gain a large z-component of momentum under the action of the radial component of the magnetic field. As a result when these particles escape the simulation box the polar angle between their momentum and the *z*-axis becomes noticeably less than $\pi/2$. However the volume of the region close to the *z*-axis scales as $\rho^2$, and the portion of particles in this region is small. The number of trapped particles is also small in comparison to number of all particles. Thus the angular spectrum is mainly determined by particles which leave the inner layer very quickly.

## 5  Threshold and structure of vacuum breakdown

The presented particle and photon emission dynamics show that two spatial regions of the m-dipole wave structure stand out. Accordingly, we decided to reveal their role in the vacuum breakdown and study each region separately. When the breakdown is studied in one region then in the other region the QED cascade is effectively switched off. Particularly, photon decay is permitted and particles experience recoil, but generated pairs are not added to the simulation. Such a method does not disturb particle or photon motion.

One may expect that vacuum breakdown should mainly develop in the inner region, where due to the larger electric field particles can gain a larger energy. However in general this is not true. We found that at the near-threshold power $P = 10$ PW if the regions are separated, the vacuum breakdown does not occur in both regions: particles escape the strong field region faster than new particles are generated as a result of the QED cascade development. We consider a cylindrical surface $\rho = 0.6\lambda$ as a boundary between the inner and outer region. For the inner layer $\rho = 0.6\lambda$ corresponds to the most distant shift of the maximum of the pair distribution from the center (see Figure 3 (a)). For the outer layer the characteristic shift of trapped particles from the second electric field node $\rho = 0.72\lambda$ is approximately $\Delta\rho = 0.1\lambda$, as was obtained above.

The absence of vacuum breakdown at the wave power $P = 10PW$ in the separated regions clearly shows that the inner and outer layers are interconnected with each other and help each other trigger the vacuum breakdown. In order to understand it several important characteristics should be used. One of the main quantum parameters, determining the possibility of the vacuum breakdown, is $\chi_{p,\gamma} \approx \eta \varepsilon_{p,\gamma} F_\perp$ [4], where indices $p$ and $\gamma$ determine particle and photon characteristics, $\eta = \hbar\omega/\varepsilon_r$, $\hbar$ is the Plank constant, $\varepsilon_{p,\gamma}$ is the particle or photon energy divided by $\varepsilon_r$, $F_\perp = \sqrt{(E + [\boldsymbol{\beta}_{p,\gamma} \times B])^2 - (\boldsymbol{\beta}_{p,\gamma} \cdot E)^2}$ is the effective transverse field, and $\boldsymbol{\beta}$ is the particle or photon velocity divided by $c$.

The particle parameter $\chi_p$ determines which fraction $\delta$ of the particle's energy it loses in an act of photon emission. On average this fraction can be determined as $\bar{\delta} = \int_0^1 \delta \frac{dW_\gamma}{d\delta} d\delta / W_\gamma$, where $W_\gamma$ is the probability of a particle to emit a photon per unit time, and $\frac{dW_\gamma}{d\delta}$ is the spectral probability density of photon emission per unit time, $W_\gamma = \int_0^1 \frac{dW_\gamma}{d\delta} d\delta$. By analogy with the effective frequency of synchrotron radiation [29] we obtained that in the range $0.17 < \chi_p < 38$ quite a proper approximation with accuracy better than 10% is $\bar{\delta} = \frac{0.21\chi_p}{2/3 + \chi_p}$.

The photon parameter $\chi_\gamma$ determines how quickly photons can decay. The expression for the probability of photon decay per unit time is $W_d = \frac{2\alpha}{3^{1.5}\eta\varepsilon_\gamma T} \int_0^1 \frac{9-u^2}{1-u^2} K_{2/3}\left(\frac{8/3\chi_\gamma}{1-u^2}\right) du$ [30]. If $\chi_\gamma \ll 1$ then the probability of photon decay per unit time is exponentially small, $W_d = \frac{1.45\alpha}{\eta\varepsilon_\gamma T}\chi_\gamma \exp\left(-\frac{8}{3\chi_\gamma}\right)$. This dependence of $W_d$ on $\chi_\gamma$ determines the necessary condition $\chi_\gamma \gtrsim 1$ for the vacuum breakdown [6] especially relevant in the case of tightly focused fields stimulating fast particle escape. According to our simulations the more precise necessary condition is $\chi_\gamma > 0.5$. This qualitative estimate can give a lower estimate for the threshold power. It does not take into account particle and photon losses or particle and photon spectra. A comprehensive study based on distribution functions in highly inhomogeneous fields is challenging to perform analytically. This study can be performed in frame of QED-PIC simulations, which we used to refine threshold powers.

In the inner region close to the electric field antinode the particle Lorentz-factor can reach $\gamma_{p,\text{in}} = a_E = 510\sqrt{P}$, the quantum parameter $\chi_{p,\text{in}} = 0.38 a_B \eta \gamma_{p,\text{in}} = 0.41P$. Here and below for estimates we consider the magnetic field amplitude as the characteristic transverse field $F_\perp$, because most of particles and photons mainly propagate close to the $z = 0$ plane (see Figure 3 (a),(b),(e),(f) and Figure 5 (b)), where the magnetic field is almost transverse to the particle momentum. Moreover, according to trajectory analysis, the magnetic field stimulates photon emission. Since in the considered range of powers $(9 < P < 50)$ $\chi_{p,\text{in}} > 1$, the maximum energy of photons is $\varepsilon_{\gamma,\text{in}}^{max} = \gamma_{p,\text{in}} = 510\sqrt{P}$, however the average photon energy is $\overline{\varepsilon_{\gamma,\text{in}}} = \gamma_{p,\text{in}}\bar{\delta}(\chi_{p,\text{in}}) = \frac{124.7P^{1.5}}{2+1.2P}$, and $\overline{\chi_{\gamma,\text{in}}} = 0.38 a_B \eta \overline{\varepsilon_{\gamma,\text{in}}} = \frac{0.1P^2}{2+1.2P}$. If this region is considered separately then the necessary condition $\overline{\chi_{\gamma,\text{in}}} > 0.5$ for the vacuum breakdown gives $P > 7.5\ PW$. However, particles escape the inner region in 0.5-0.7T and characteristic escape time of photons is approximately $l_F/c = 0.4T$ ($l_F$ is the characteristic scale of the field structure). This is the main reason why the vacuum breakdown threshold power in the separated inner region is 1.6 times greater than given by the estimate above and is approximately 12.5 PW according to QED-PIC simulations. The threshold is determined as in the section 3.

In the outer region close to the electric field antinode the particle Lorentz-factor can reach $\gamma_{p,\text{out}} = 0.4 a_E = 200\sqrt{P}$, the quantum parameter $\chi_{p,\text{out}} = 0.34 a_B \eta \gamma_{p,\text{out}} = 0.15P$. Since $\chi_{p,\text{in}} > 1$ within $9 < P < 50$, the maximum energy of photons is $\varepsilon_{\gamma,\text{out}}^{max} = \gamma_{p,\text{out}} = 510\sqrt{P}$. The average photon energy is $\overline{\varepsilon_{\gamma,\text{out}}} = \gamma_{p,\text{out}} \bar{\delta}(\chi_{p,\text{out}}) = \frac{18 P^{1.5}}{2 + 0.44P}$, and the average quantum parameter is $\overline{\chi_{\gamma,\text{out}}} = 0.34 a_B \eta \overline{\varepsilon_{\gamma,\text{out}}} = \frac{0.013 P^2}{2 + 0.44P}$. Thus in the outer region the necessary condition dictates that the wave power should be $P > 21\ PW$ to trigger vacuum breakdown. In this region particles need a long time of about 3T to leave the trapping region, however photons escape the outer region within 0.25T. According to the QED-PIC simulations, escape of particles and photons increases the threshold power in the separated outer region up to 23 PW. The difference between the analytically estimated threshold and the threshold obtained in frame of QED-PIC simulations is smaller in the outer region because there the particle escape rate is much slower.

Thus if regions were separated then the vacuum breakdown threshold power would be 12.5PW and would refer to the inner region. However, the interconnection of outer and inner regions decreases the threshold power down to 10PW. Since particles are not trapped in the inner region and quickly leave this region, in order to trigger vacuum breakdown photons from the outer region must decay in the inner region. The energy of these photons is $\overline{\varepsilon_{\gamma,\text{out}}}$, and propagating to the center they can reach the maximum magnetic field $a_B$ where their quantum parameter $\overline{\chi_{o-1}} = a_B \eta \overline{\varepsilon_{\gamma,\text{out}}} = \frac{0.038 P^2}{2+0.44P}$. From the lower estimate $\overline{\chi_{o-1}} > 0.5$ it follows that the vacuum breakdown is possible when $P > 8.8 PW$, which is very close to the vacuum breakdown threshold $P_{th} \approx 10 PW$ given by QED-PIC simulations. Note that photons from the inner region can also decay in the outer region. In this case the photon quantum parameter is $\overline{\chi_{1-o}} = 0.34 a_B \eta \overline{\varepsilon_{\gamma,\text{in}}} = \frac{0.089 P^2}{2+1.22P} > 0.5$ when $P > 8.2 PW$.

These estimates show that the interconnection of the two regions is as follows. Photons from the outer regions can decay in the inner region and compensate fast particle escape. In turn, generated particles leaving the inner region can be trapped in the outer region. Also, photons generated in the inner region can decay in the outer region. Thus, during vacuum breakdown, pair production occurs in both outer and inner regions, and at near threshold wave powers inner and outer layers are equally pronounced. According to the QED-PIC simulations at larger powers $P > 30 PW$, when generation of new photons and particles becomes faster than particle and photon transitions from one region to another, the inner region becomes more prominent because there particles and photons achieve the highest quantum parameters $\chi_{p,\text{in}}$ and $\overline{\chi_{\gamma,\text{in}}}$.

To illustrate our conclusions the time structure of the vacuum breakdown is shown in Figure 6 for the wave power 10PW. This near-threshold wave power is chosen to make the periodic nature of processes more visual. Figure 6 (a), (b) shows that the surface $\rho = 0.6\lambda$ determined from the particle density distribution is suitable as a boundary between the outer and inner regions. The photon and pair productions in both regions are modulated in time and when magnetic field is negligible ($t = 0.5kT, k \in \mathbb{Z}$) then pair and photon productions are minimal.

Also in the vicinity of the magnetic field node $\rho = 0.44\lambda$ the photon production and photon decay are suppressed. These two facts emphasize the role of the magnetic field as the main component of the effective transverse field $F_\perp$.

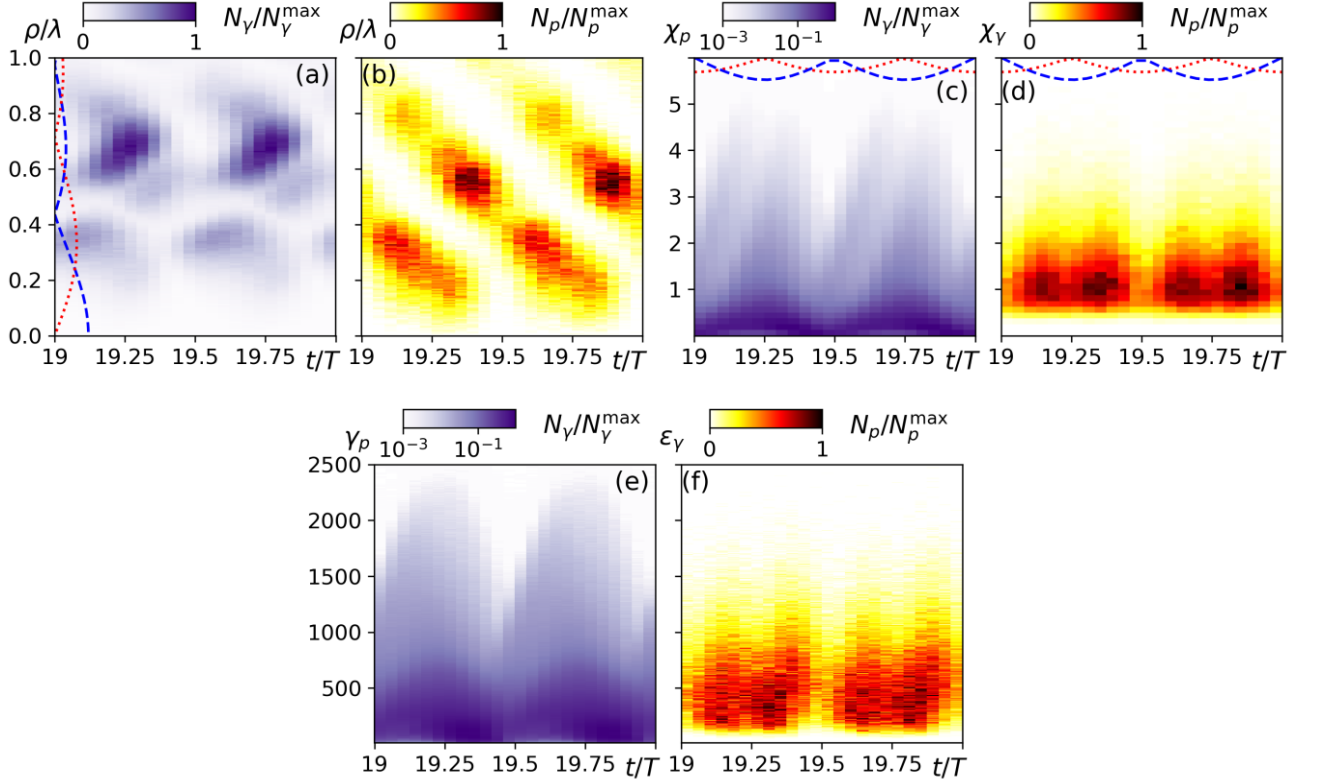

**Figure 6.** Number of events of photon $N_\gamma$ and pair $N_p$ production as function of time and (a),(b) radial coordinate $\rho$, as function of (c) parent electron(positron) quantum parameter $\chi_p$ and (d) photon quantum parameter $\chi_\gamma$, and (e) parent electron(positron) Lorentz-factor $\gamma_p$ and (f) photon energy $\varepsilon_\gamma$, normalized to the electron rest energy.

For $P = 10 PW$ the maximum quantum parameter of particles $\chi_{p,\max} \approx 5$ (see Figure 6 (c)) and is consistent with $\chi_{p,\text{in}} \approx 4$, which is the maximum value among estimates of particle quantum parameters $\chi_p$ in inner and outer regions. Figure 6 (d) clearly demonstrates that there are almost no acts of photon decays if $\chi_{\gamma,\min} < 0.5$, so the chosen above necessary condition for vacuum breakdown $\chi_\gamma > 0.5$ is reasonable. The majority of events of photon decay occur with $\chi_\gamma \approx 1$, because for larger $\chi_\gamma$ the number of photons significantly decreases (see Figure 6 (d)) and for smaller $\chi_\gamma$ the probability of photon decay becomes exponentially small. The maximum photon quantum parameter is approximately the same as for particles and corresponds to photons which take almost all kinetic energy from particles.

The particle Lorentz-factor varies in range $[1; 2300]$ (see Figure 6 (e)), which is qualitatively similar to the range $[1; \gamma_{p,\text{in}} \approx 1600]$ estimated above. The energy distribution of photons (see Figure 6 (d)) is within a smaller range. The majority of photon decays happens in range $\varepsilon_\gamma \in [150, 750]$, and the maximal photon energy corresponds to the maximal Lorentz-factor of the parent particle. A qualitative estimate of the average photon energy $\overline{\varepsilon_{\gamma,\text{in}}}$ in the inner layer is exactly at the middle of this range, but the estimated average energy $\overline{\varepsilon_{\gamma,\text{out}}}$ in the outer layer is underestimated by $\approx 1.6$ times.

According to Figure 6 (d) photons can decay if $\varepsilon_\gamma > 110$. This photon energy corresponds to the minimal quantum parameter ($\chi_{\gamma,\min} = 0.5$) sufficient for photon decay. In the strongest transverse field $F_\perp \approx a_B$ the qualitative estimate of the minimal energy is $\varepsilon_{\gamma,\min} = 0.5/(\eta a_B) \approx 75$ and it correlates with the corresponding result of the numerical simulation.

Finally, the qualitative analysis and numerical simulations of the vacuum breakdown is presented. They explain the interconnection of inner and outer $e^-$-$e^+$ layers and show temporal dynamics of the breakdown. Derived qualitative estimates differ from the values obtained as a result of the numerical simulation by a factor no larger than 2.

# 6  Experimental approach

From the experimental point of view it is crucially important to understand attainability of the vacuum breakdown in fields of the m-dipole wave. The important question is the creation of this field configuration. The m-dipole wave can be mimicked similarly to the e-dipole wave with the help of a number of linearly polarized beams [23, 24] but with orthogonal polarization. The threshold power of the vacuum breakdown of 10PW demands less than 1 PW per tightly focused laser beam in the case of a 12-beam configuration. Nowadays petawatt lasers are widespread [1], so in principle there is a possibility of experimental realization of the m-dipole wave with power in the 10-PW range.

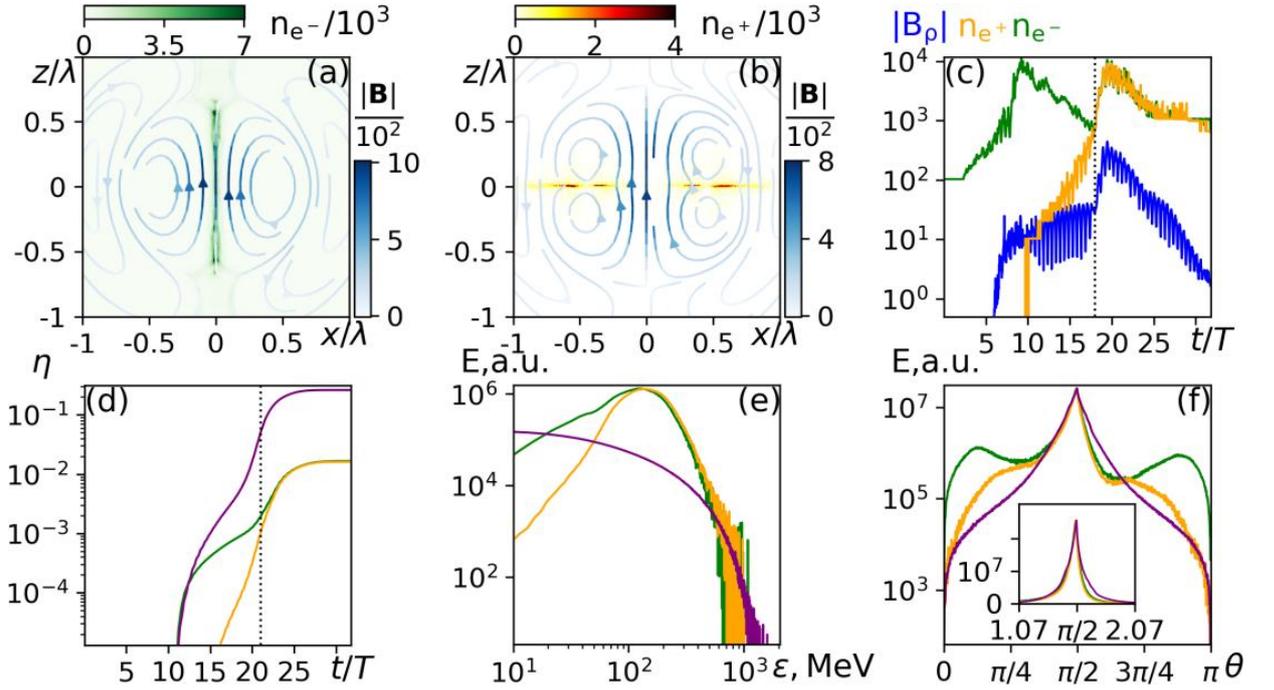

**Figure 7. Vacuum breakdown triggered by interaction of 30 fs m-dipole pulse of 20 PW peak power with nanowire target with density $100 n_c$ and radius $0.25\lambda$ located in focus of laser pulse. (a), (b) Comparison of magnetic field lines and electron-positron distribution before ($t = 10.1T$) vacuum breakdown and after ($t = 18.9T$) it at the self-compression stage. (c) Time evolution of maximal values of radial component of the magnetic field and maximal densities of electrons and positrons. (d) Efficiency of energy transformation from the laser energy to energy of particles (green line for electrons and orange line for positrons) and photons (purple line for photons) leaving the focal region. Energies are measured for particles crossing a sphere with radius $3\lambda$. Dotted lines in (c) and (d) correspond to beginning of nonlinear stage. (e) Energy spectra and (f) angular distribution of escaping particles and photons measured as in figure (d). Polar angle $\theta$ is counted from the z-axis.**

Pulse durations in case of such strong laser beams is expected at the level of 10 laser periods and the question arises whether this duration is enough for the vacuum breakdown to occur and for reaching the nonlinear stage. To answer this question it is necessary to consider a certain target. Without detailed selection of a target we choose a nanowire target with a radius of $0.25\lambda$ and electron and proton densities of $100n_{\text{cr}}$. First, simulations with a half-infinite wave described in the previous section showed strong plasma compression by the incident pulse in the direction transverse to the *z* axis. The nanowire target placed at the point of beam focusing allows shortening of the stage of initial compression. Second, a target with a solid density leads to shortening of the avalanche stage and makes it possible to reach the nonlinear stage of plasma self-compression.

We performed a series of QED PIC-code simulations modeling an interaction of the nanowire target with the ideally focused m-dipole pulse with a $\sin^2$-envelope. Its duration was 30 fs (full width at half-maximum of intensity). The time step was $dt = T/300$, duration of simualtion was 37T. The simulation box was $3\lambda \times 3\lambda \times 3\lambda$ with $768 \times 768 \times 768$ cells along *x*, *y* and *z* axes. The initial number of macroparticles of each type and the threshold number of particles for resampling [31] were $1.2 \times 10^7$ and $4.8 \times 10^7$, respectively. Initially one macroparticle corresponds to approximately $10^4$ real physical particles.

According to simulations even a 10 PW 30 fs m-dipole pulse is enough for the vacuum breakdown and achievement of the avalanche stage. This happens because the target density is quite large and the presence of protons impedes electron escape and consequently the threshold power is decreased; a similar effect was observed in the fields of the e-dipole wave [24] and in colliding laser pulses [21]. However, to achieve the nonlinear stage a minimum of 17PW power is needed. For comparison we once again present results of simulation for a 20 PW 30 fs pulse of the m-dipole wave (see Figure 7).

The laser pulse compresses electrons of the target towards the *z*-axis into a narrow string and their density reaches overcritical values of about $10^4 n_{\text{cr}}$ at $t \approx 10T$, the standing wave is formed and production of pair plasma starts (see Figure 7(a),(c)). After this the electron string decays due to field inhomogeneity, so the maximal electron density decreases. At the same time a structure of pair plasma transverse to the *z*-axis appears, so the positron density increases. Maximal $B_\rho$ starts to grow following the pulse envelope. During the avalanche stage transformation of laser energy to the energy of photons is substantial while for particles it is not the case. The avalanche stage is followed by the nonlinear stage starting at $t \approx 18T$, and at this time photons, electrons and positrons possess about 5%, 0.2% and 0.1% of total pulse energy, respectively (see Figure 7(d)). Since the energy of particles and photons is measured at distance $3\lambda$ from the central point, the registered particles and photons correspond to the plasma state approximately 3T earlier. So in Figure 7 (c) the transition to the nonlinear stage can be observed at $t \approx 21T$.

In one laser period after the beginning of the nonlinear stage quasi-neutral electron-positron plasma density increases by a factor of 15 up to $\sim 10^4 n_{\text{cr}}$, $B_\rho$ increases by a factor of 13 up to ~450, thus a current instability develops. Plasma is compressed into layers with thickness 7 nm (the cell size of the simulation box) and magnetic field lines reconnect (see Figure 7(b)) and these lines form x-point-like magnetic structures [32]. The maximal compression lasts about $2.5T$. After this, fields of the incident pulse decrease and the excess of particles for the instantaneous wave power leaves the focal region causing a decrease of plasma density. During the nonlinear stage 21% of laser energy is transformed to photon energy and particles amount to approximately 1.5% of laser energy (see Figure 7(d)).

Since the amount of particles is smaller at the avalanche stage in comparison with the nonlinear stage, only the second stage determines energy and angular spectra of particles and photons (see Figure 7(e),(f)). The width of angular distributions of particles and photons is approximately 6 degrees. So, the simulation confirms a possibility to use angular spectra as a signature of a vacuum breakdown stage.

Also, we should note that the tails of the angular distribution below 10% of the distribution's peak lose symmetry relative to $\theta = \pi/2$, different maxima are formed (see Figure 7(f)). According to simulations it is related to oscillations of the compressed layer at a later nonlinear stage, which are the subject of further research.

Thus, the vacuum breakdown in the m-dipole wave looks achievable in principle based on existing laser technologies. 10 PW power of the m-dipole 30 fs pulse is sufficient to trigger vacuum breakdown and create the distinguishable electron-positron plasma structures. In the case of power greater than 17 PW it is possible to observe the nonlinear stage, during which a current instability develops, compressing the generated quasi-neutral electron-positron plasma to a layer with thickness less than 10nm and density greater than $10^{25}$ cm$^{-3}$. Moreover this stage can be distinguished from the avalanche stage by the energy and angular spectra of particles and photons. Particles and photons escape the focal region at polar angle $\pi/2$ with angular spread of about 10 degrees, in contrast to ~1 degree at the avalanche stage. Also particle and photon energies are several tens of percents less at the nonlinear stage than at the avalanche stage. Laser energy is transformed to particle energy with efficiency of a few percent and to gamma radiation with efficiency of several tens of percents. Energy spectra of particles and photons stretch to GeV energies.

## 7 Conclusion

We have studied the vacuum breakdown in laser fields in the form of the converging m-dipole wave, which maximizes the magnetic field, in the power range of 9 PW to 50 PW with the help of QED-PIC simulations. The main attention has been paid to the avalanche stage of QED cascade development when the generated e$^-$-e$^+$ plasma has a low density insufficient to influence back on the driving laser wave. The space-time distribution of plasma is similar to oscillating concentric cylindrical layers and is determined not only by the space-time structure of pair production but also by particle motion. In this power range newly born particles can quickly escape the focal region or be trapped in the electric field node (outer) region moving in the *normal radiative trapping* regime. Trapped particles abundantly emit photons in the direction towards the focus and due to photon decay the region around the first electric field antinode (inner region) is also populated by e$^-$ and e$^+$. The nonlocality of the QED cascade is of great importance and coupling of these two regions determines the threshold power of the vacuum breakdown which is approximately 10 PW. However, at greater powers this coupling becomes less significant and the pair plasma layers within the inner region are more pronounced. Although the m-dipole wave is not as optimal to trigger the vacuum breakdown as the e-dipole wave (for which threshold power is about 7 PW), nevertheless the threshold power and the avalanche growth rate are close in these modes. Since the main instrument was 3D PIC simulation we have also made a full modeling of nonlinear laser plasma interaction and showed that relativistically dense $e^-$-$e^+$ plasma can be produced in the m-dipole wave with laser-plasma structures different as compared to the case of the e-dipole wave. Results of this study will be published elsewhere.

**Acknowledgements**

The reported study was funded by RFBR and ROSATOM according to the research project № 20-21-00095.